# Supported lipid membranes with designed geometry

Melissa Rinaldin, Sebastiaan L. D. ten Haaf, Ernst J. Vegter,
Casper van der Wel, Piermarco Fonda, Luca Giomi, and Daniela J. Kraft*

Dr. M. Rinaldin: Leiden Institute of Physics, University of Leiden, 2300 RA Leiden, The Netherlands, now Max Planck Institute of Molecular Cell Biology and Genetics, Pfotenhauerstraße 108, 01307 Dresden, Germany.
S. L. D. ten Haaf: Leiden Institute of Physics, University of Leiden, 2300 RA Leiden, The Netherlands.
E. J. Vegter: Leiden Institute of Physics, University of Leiden, 2300 RA Leiden, The Netherlands.
Dr. C. van der Wel: Leiden Institute of Physics, University of Leiden, 2300 RA Leiden, The Netherlands.
Dr. P. Fonda: Instituut-Lorentz, Universiteit Leiden, Leiden, 2300 RA, The Netherlands, now Max Planck Institute of Colloids and Interfaces, Potsdam, 14476, Germany
Dr. L. Giomi: Instituut-Lorentz, Universiteit Leiden, Leiden, 2300 RA, The Netherlands.
Dr. D. J. Kraft: Leiden Institute of Physics, University of Leiden, 2300 RA Leiden, The Netherlands, kraft@physics.leidenuniv.nl.

**Abstract:** The membrane curvature of cells and intracellular compartments continuously adapts to enable cells to perform vital functions, from cell division to signal trafficking. Understanding how membrane geometry affects these processes *in vivo* is challenging because of the membrane complexity as well as the short time and small length scales involved. By contrast, *in vitro* model membranes with engineered curvature provide a versatile platform for this investigation and applications to biosensing and biocomputing. However, a general route to the fabrication of lipid membranes with prescribed curvature and high spatial resolution is still missing. Here, we present a strategy that overcomes these challenges and achieve lipid membranes with designed shape by combining 3D micro-printing and replica-molding lithography to create scaffolds with virtually any geometry and high spatial resolution. The resulting supported lipid membranes are homogeneous, fluid, and can form chemically distinct lipid domains. These features are essential for understanding curvature-dependent cellular processes and developing programmable bio-interfaces for living cells and nanostructures.

**Main text:**

Controlling the geometry of *in vitro* lipid membranes is critical for advancements in cellular biology and bioengineering [1] [2] [3]. Current approaches to control the shape of model membranes consist of manipulating lipid vesicles *in situ* [4] [5] or creating supported lipid bilayers [3]. These techniques, however, give access to a limited range of shapes only, which makes them unsuitable for investigating curvature-dependent cellular processes, such as intracellular signaling [6] [7], applications such as live-cell manipulation [8] [9] [10] [11] and DNA origami circuit construction [12] [13] [14]. To overcome these limitations, we here present a strategy based on membrane-supporting microstructures created with 3D object rational design, direct laser writing and subsequent replica-molding with polydimethylsiloxane (PDMS). The versatile workflow specific to 3D micro-printing allows to create lipid membranes of virtually any geometry. The PDMS used in the replica-molding step enables the crucial homogeneity and fluidity of the membrane. Moreover, the elastic nature of PDMS mimics the flexible actin cellular cortex and it can be used to apply active deformations to the lipid membrane [15]. Furthermore, a tight control of membrane curvature and tunable softness of supporting PDMS substrates are both key features for a faithful representation of the cell interface, required for example in immunological studies with T cells [16]. Our method allows programming membrane curvature with extensive design freedom and high spatial resolution, and hence could provide a versatile platform for understanding cell biology processes associated with distinct membrane geometries and for engineering bio-interfaces.

We illustrate our strategy for creating supported lipid membranes with designed shape in **Figure 1a**. First, we designed the desired support structure by creating 3D molds of the inverted shape using AutoCAD Inventor. We fabricated the molds by 3D micro-printing via two-photon polymerization using a commercially available printer (**Figure 1a**, step 1). To achieve high spatial resolution, arrays of the molds were printed onto silicon wafers using the photoresist IP-S and 63x oil-immersion objective (Zeiss, NA=1.48). A femtosecond laser traced the mold shape through galvanic mirrors, and a mechanical stage controlled with a piezoelectric actuator allowed displacing the substrate to create larger structures. The accessible size range, from hundreds of nanometers to millimeters, and spatial accuracy down to 100 nm can only be obtained through 3D direct laser writing [17]. The molds were finalized by developing the as-printed resist in a solution of propyleneglycolmethyletheracetate (PGMA) under UV light. Subsequently, to create the scaffold, a solution of PDMS pre-polymer was poured over the molds and cured at a temperature of 100 °C for 2 hours. The obtained polymerized PDMS has a Young's modulus of 2.5 ± 0.1 MPa [18] (**Figure 1a**, step 2). Finally, the polymerized PDMS was peeled from the molds to obtain the substrate patterned with an array of the desired support structure, for example Gaussian bumps (**Figure 1a**, step 3). While the mold has a resolution of 100 nm set by the printing accuracy, the PDMS structures present a smoother surface which reduces its surface area. This replica molding with PDMS was a critical step to fabricate substrates suitable for lipid functionalization. Indeed, lipid coating of substrates made of IP-S or Ormocomp photoresists led to homogeneous but partially fluid membranes, with significantly slower diffusion constants ($D_{FRAP, IP-S}$= 0.2 ± 2.0 µm$^2$/s and $D_{FRAP, Ormocomp}$ = 0.40 ± 0.01 µm$^2$/s) and recovery percentages of 28% and 60%, respectively, as measured by fluorescent recovery after photobleaching (FRAP). (**Supporting Information, Figure S1 and Table S1**). Finally, we coated the PDMS supports with a lipid membrane via vesicle deposition. To prevent monolayer formation [19], we hydrophilized the PDMS surface by thorough cleaning with sequential sonication in acetone, ethanol, and MilliQ water and UV/ozone treatment. Then, we positioned the PDMS microstructures in a flow cell where a solution of small unilamellar vesicles (SUVs) made of a mixture of lipids consisting of 94.8% DOPC, 0.2% DOPE Rhodamine, and 5% DOPE-PEG (2000) mole ratio was flown in (**Figure1b**, step 4).

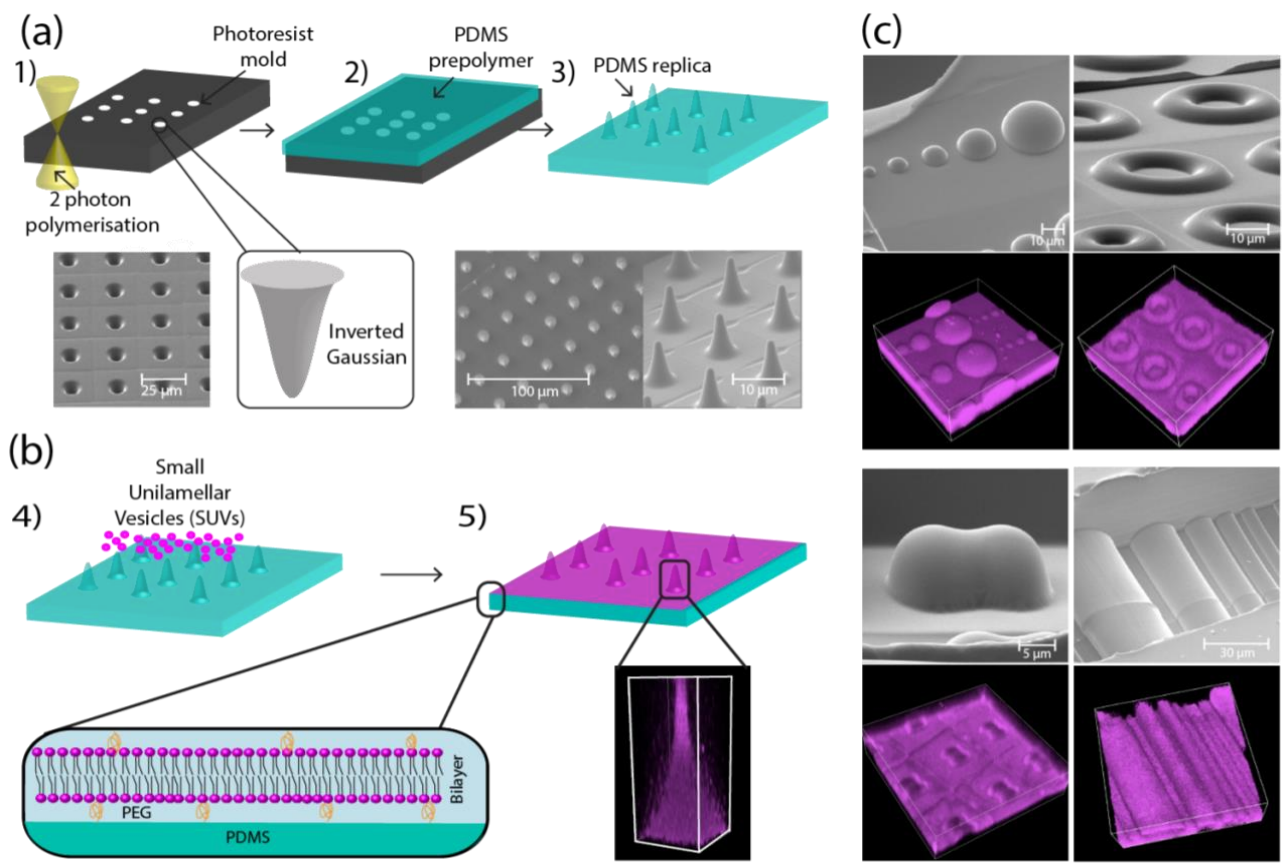

**Figure 1: Fabrication of supported lipid membranes with designed shape. (a) Microstructures fabrication**. Step 1) Molds of Gaussian bumps are fabricated by two-photon polymerization. Step 2) PDMS pre-polymer is poured over the developed microprinted molds and cured. Step 3) The resulting microstructures are peeled from the molds. Scanning electron microscopy (SEM) images of the PDMS Gaussian bumps (right) and their mold counterparts (left) are shown in the bottom row. **(b) Membrane formation**. Step 4-5) The microstructures are functionalized with a lipid membrane by deposition of small unilamellar vesicles (SUVs). The lipid composition is 94.8% DOPC, 0.2% DOPE Rhodamine, and 5% DOPE-PEG(2000) mole ratio. On the bottom-left, a close-up schematic view of the membrane supported on a PDMS substrate is shown. PEGylated lipids DOPE-PEG (2000) (orange) are used to increase the water layer between the PDMS substrate and the membrane and therefore increase the fluidity of the membrane (magenta). On the bottom-right, a 3D fluorescence microscopy image of one of the Gaussian bumps is reported. **(c) Membranes with designed shape**. SEM images of PDMS microstructures of with half-spherical and half-toroidal (top), and half-dumbbell and half-cylindrical shape (bottom) along with fluorescence volume reconstruction of the membranes

The use of DOPE-PEG (2000) increases the distance between lipids and substrate and hence preserves membrane fluidity, while DOPE Rhodamine allows for imaging the membrane using confocal microscopy. After one hour of incubation, the PDMS substrate was fully covered by the membrane. The supported lipid membrane follows the geometry of the mold perfectly within the resolution of confocal microscopy, and likely even down to the nanoscale precision [20] (**Figure 1b**, step 5).

The versatility of rational 3D micro-printing design allows one to straightforwardly create a library of shapes and mimic the myriad of conformations taken on by cellular membranes, ranging from protrusions in filopodia and cilia as well as pits in clathrin-mediated endocytosis [8]. In **Figure 1c**, we report examples of hemispherical, half-toroidal, -dumbbell, -cylindrical microstructures, with and without the lipid coating. This selection of shapes allows us to manipulate the local curvatures of the membrane: the hemisphere has homogeneous mean and Gaussian curvatures. The half-torus has a varying mean and Gaussian curvature (positive inside and negative outside), while cylinders have homogeneous mean curvature and vanishing Gaussian curvature. Half-dumbbells contain valley-like features which are reminiscent of size-controlled necks in supported lipid vesicles (SLVs), which have been used to measure Gaussian rigidity difference in multicomponent membranes [21]. Furthermore, similar minimal neck-like shapes could be used for more precise measurements of the saddle splay modulus via flickering interface analysis [22].

To functionalize membranes with biomolecules and nanostructures [10], the supported membrane must be homogeneous and fluid [23]. We tested whether our supported membranes feature homogeneity by examining the fluorescence intensity of the bilayer on 3x3 array of 10 μm-diameter hemispheres. We found that the intensity is homogenous across the membrane and independent of the curvature of the substrate (**Figure 2a**). Moreover, the fluorescence intensity along a line taken on the top of a hemisphere randomly fluctuates around its mean value, indicating that the membrane is homogeneous (**Figure 2b-c**). To measure membrane fluidity, we performed FRAP experiments (**Figure 2d-e**). We simultaneously bleached a circular area around the top of 8 hemispheres, and observed complete recovery of the intensity, see **Figure 2d**. Indeed, the normalized intensity for each sphere which includes a correction for imaging photobleaching, follows roughly the same exponential curve, with an average half time of $\tau_{1/2}$ = 4.8 s and average mobile lipid fraction 93% (**Figure 2e**). See **Supporting Information, Table S2** for details on the fit parameter.

The combination of designed geometry, preserved homogeneity and fluidity makes our protocol a powerful platform for quantitative studies on geometry-dependent effects on model membranes. To demonstrate this, we quantify the impact of curvature on the FRAP diffusion coefficient [24] [25] [26] [27]. We performed FRAP experiments at the top of hemispheres of five different radii and measured the FRAP diffusion coefficient ($D_{FRAP}$). To account for the fact that the bleached area is a spherical cap, we modified the expression of $D_{FRAP}$ given by Axerold *et al.* [28] by using the surface area of a spherical cap instead of a flat disk (see **Supporting Information, Section III**). We observe in **Figure 2f** that $D_{FRAP}$ depends on the substrate shape, and in particular that $D_{FRAP}$ is inversely proportional to the square of the radius of the half-spherical substrate. While the actual diffusion coefficient of the lipids is not affected by the curvature, the radius dependence of $D_{FRAP}$ captures the collective effect of the half-spherical geometry on the mean square displacement (MSD) of the single lipids. Indeed, by expressing the MSD of particles diffusing on a sphere [29] in terms of a "geometry-dependent" diffusion coefficient, the radius dependence of D$_{FRAP}$ found in **Figure 2f** and the actual diffusion coefficient can be recovered (see **Supporting Information, Section IV**).

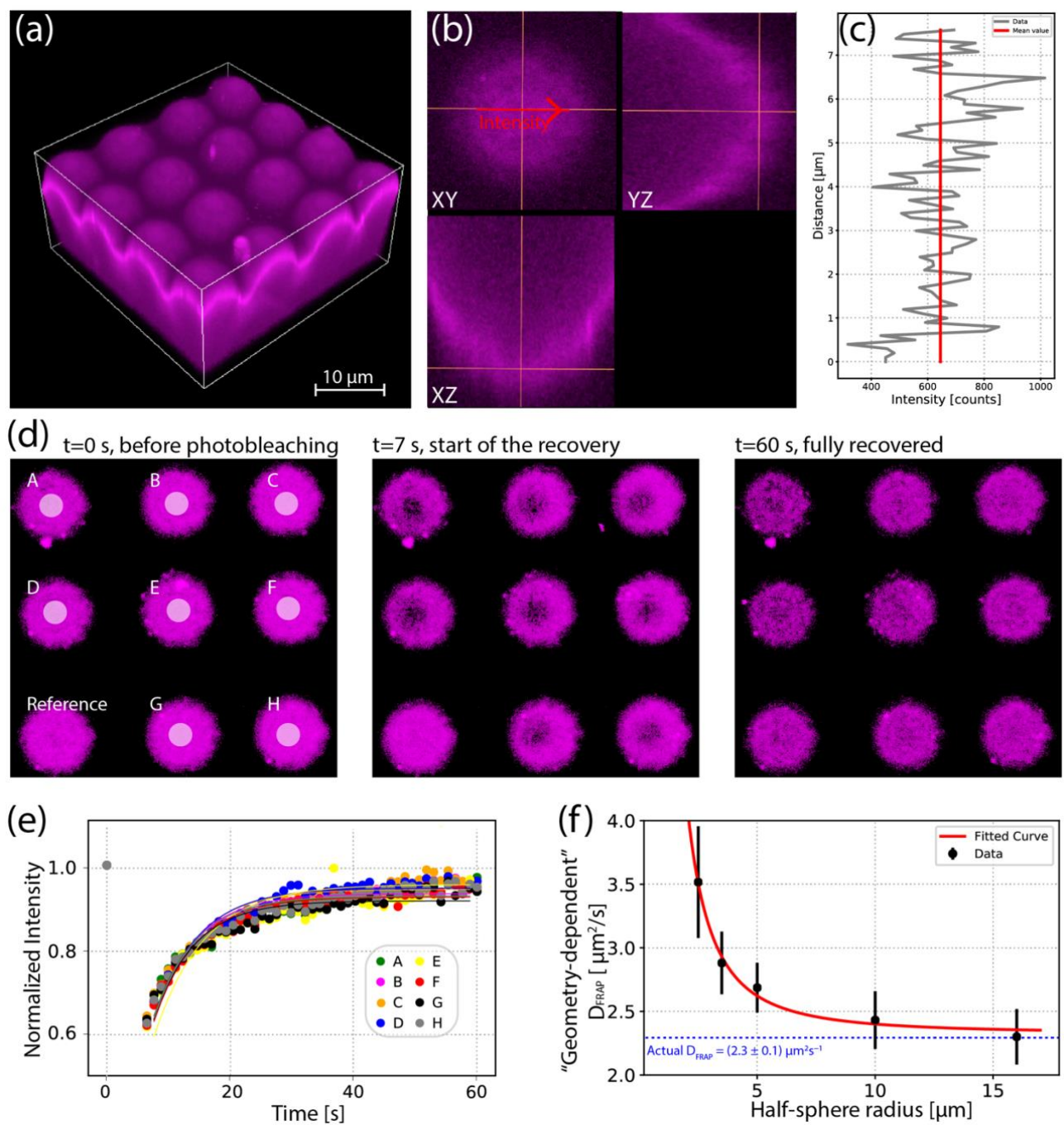

**Figure 2: Membrane homogeneity and fluidity. (a)** 3D reconstruction of a membrane on half-spherical microstructures of 5 μm radius. **(b)** XY, YZ, and XZ projections of the membrane on the top of one hemisphere. The intensity of the fluorescent signal along the grey axis is plotted in **(c)** which shows the intensity fluctuating around an average value, indicating that the membrane is homogeneous. **(d)** FRAP experiments on the top of eight half-spherical membranes. The bleaching areas (white disks) are labelled with letters from A to H. From left to right, the stages before and after bleaching, as well as at the end of the recovery are reported. **(e)** Experimental data (cirlces) and fits (lines) of the FRAP experiments. The intensities recover almost completely. **(f)** FRAP "geometry dependent" diffusion coefficient as a function of the radius of the hemispheres, with a least-square using $D_{FRAP}=A + B/R^2$, where A = (2.3 ± 0.1) μm$^2$s$^{-1}$ which is also the value of the actual diffusion coefficient and B= (7.4 ± 0.2 μm$^4$s$^{-1}$).

Homogenous fluid lipid membranes scaffolded onto designed surfaces offer a compelling opportunity to precisely engineer the formation of local heterogeneities via liquid-liquid phase separation. For specific temperatures and lipid compositions, artificial lipid membranes undergo liquid-liquid phase separation from one homogenous liquid phase into two distinct liquid phases: a more rigid, liquid-ordered phase (LO), and a softer, liquid-disordered (LD), phase [30]. In the presence of curvature, the LO and LD domains organize into specific patterns dictated by the curvature of the underlying substrate [31] [21] [32] [33]. We induced lipid phase separation in our set-up by using a lipid mixture containing cholesterol, sphingomyelin, and POPC in mole ratio equal to 25:50:25. Moreover, the SUV formation and lipid coating were performed at 70 $°C$, *i.e.* well above the critical transition temperature, which is expected to be between 35 $°C$ and 45 $°C$ [30] . After the lipid coating, the membrane was cooled to 25 $°C$ at a rate of 0.04 $°C min^{-1}$. In **Figure 3a** we show that we achieved lateral phase separation on half-spherical microstructures of varying radius. The LD phase was labelled in magenta using 0.2% mol DOPE-Rhodamine and the circular LO domains were unlabeled and hence appear dark (**Figure 3a**). While on the flat parts of the membrane the LO domains are spatially randomly distributed and feature similar sizes, on the hemispheres the occurrence and size of the LO domains are influenced by the substrate curvature. This can be seen in **Figure 3c**, where we report top views of half-spherical membranes as sketched in **Figure 3b.** On the smallest hemisphere of radius 2.5 µm there are no observable LO domains. By increasing the radius, more numerous and larger LO domains are present on the half spheres because of the larger surface available and the preference of LO domains for less curved regions [31] [31] [33]. The presence of multiple circular domains indicates that the system has locally, and not globally reached equilibrium. The resulting inhomogeneous spatial distribution of lipid domains suggests that we can use this technique for fabricating curved substrates with chemically distinct domains which can be functionalized at will with biomolecules and nanoparticles, for example through biotin-streptavidin binding to (un)saturated lipids.

In conclusion, we presented a novel approach to fabricating a plethora of homogenous and fluid curved lipid membranes in a precise and programmable fashion. We exploited the rational design capabilities of 3D micro-printing to obtain specific anisotropic structures and used the biocompatibility of PDMS to achieve homogeneous and fluid membranes which can feature chemically distinct domains upon using a ternary lipid mixture. Our membranes with designed shapes are a powerful platform for unravelling curvature-dependent biological processes in membranes, such as sorting of proteins and lipids in signal trafficking. Moreover, our strategy opens up new possibilities in bio-engineering and material science. Specifically, the development of synthetic platforms to impose curvature on living cells for mechanobiology [8] [9] [10] [11], immunology experiments [16], and bottom-up construction of complex nanostructures of DNA-origami building blocks [12] [13] [14].

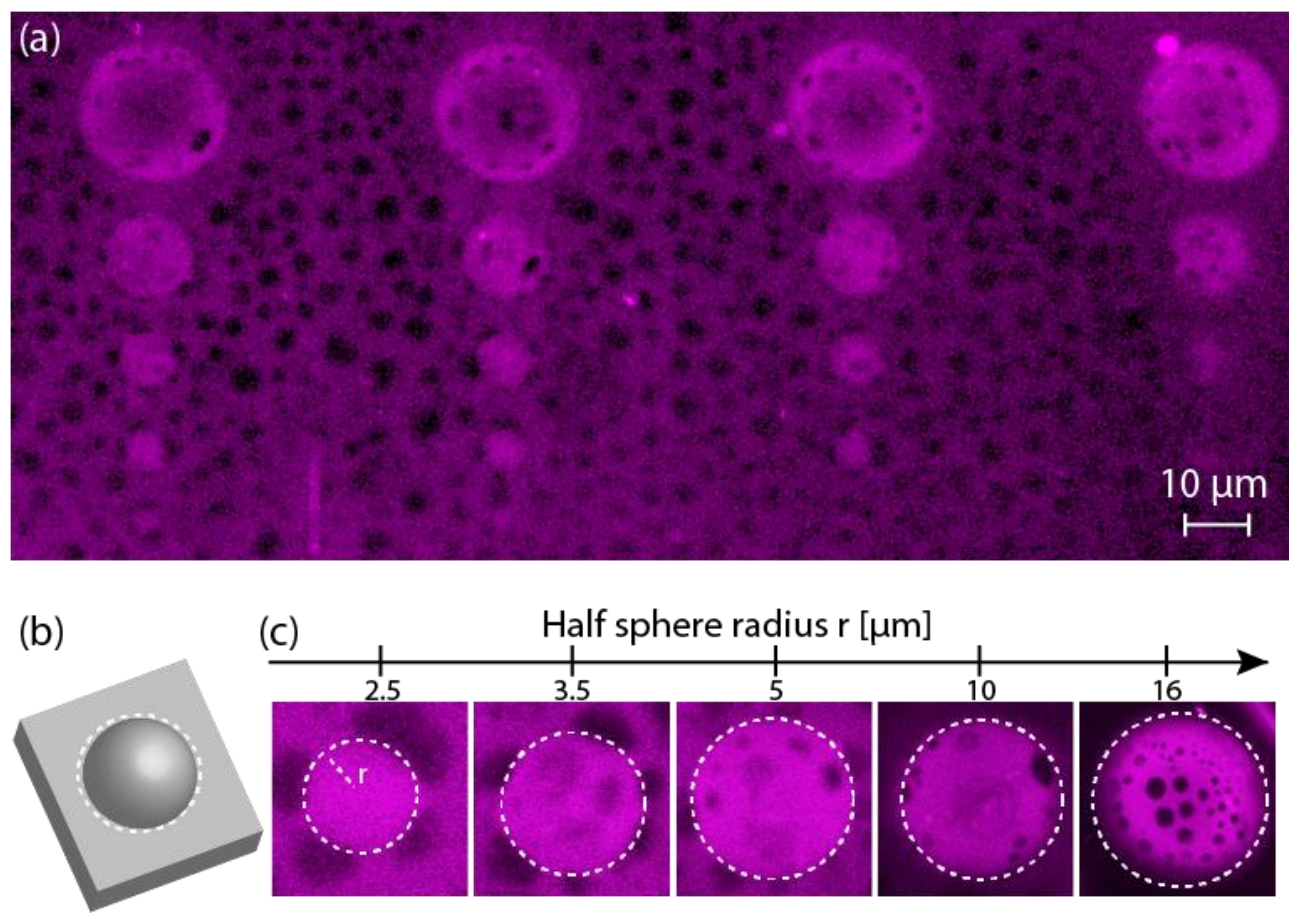

**Figure 3: Phase-separated membrane on half-spherical microstructures of varying radius. (a)** Fluorescence microscopy image of a phase-separated membrane on half-spherical microstructures. LD phase is labelled by DOPE-Rhodamine (magenta). Circular LO domains are visible both on the flat substrate and the half-spherical structures. **(b)** Schematic view of the half spheres from the top, as shown in **(c)**. Top views of the phase-separated membranes on hemispheres with 2.5, 3.5, 5, 10, and 16 μm radius. The LD phase is shown in magenta. The boundary of the hemispheres is shown by dotted white circles. For the smallest radius, LO domains are depleted from the hemisphere. As the radius increases, or the curvature decreases, the amount of LO domains on the hemispheres increases indicating a preference for the LO domains for less curved regions.

**Acknowledgements**

This work was supported by the Netherlands Organization for Scientific Research (NWO/OCW), as part of the Frontiers of Nanoscience program. We thank Joseph Salaris for preliminary experiments, Jeroen Mesman for the fabrication of the flow cell for lipid functionalization, and Rachel Doherty for the electron microscopy imaging.

**Experimental**

**Reagents:** The lipids (Δ9-Cis)1,2-dioleoyl-sn-glycero-3-phos-phocholine (DOPC), 1-palmitoyl-2-oleoyl-sn-glycero-3-phosphocholine (POPC), porcine brain sphingomyelin(BSM), ovine wool cholesterol (Chol), 1,2-dioleoyl-sn-gly-cero-3-phosphoethanolamine-N-[methoxy(polyethyleneglycol)-2000] (DOPE-PEG(2000)), and L-α-Phosphatidyl-ethanolamine-N-

(DOPE lissamine rhodamine B sulfonyl) were purchased from Avanti Polar Lipids and stored at −20°C. Propylene glycol monomethyl ether acetate (PGMA, ≥99.5 %) and trichloro(1H,1H,2H,2H-perfluorooctyl) silane (≥97 %) were purchased from Sigma-Aldrich. 2-Propanol (≥99.6 %) was purchased from Honeywell. Silcon wafers of 10 cm diameter, 1 Ωcm resistivity, and 525 μm thickness were purchased from Siegert Wafer. IP-S photoresist was purchased by Nanoscribe GmbH and Ormocomp by Microresist. Polymethyl siloxane (PDMS) was purchased from Dow Corning.

**Fabrication of solid substrates:** Microstructures for membrane formation were prepared by PDMS replica-molding from 3D printed molds. The 3D structure of the molds was designed using the software Autodesk Inventor (Autodesk) and processed with the program Describe (Nanoscribe GmbH) to obtain a mesh of points suitable for micro-printing. Subsequently, the molds were fabricated by two-photon lithography using a commercial 3D microprinter (Nanoscribe GmbH, Photonics Professional GT). The photoresist (IP-S, Nanoscribe GmbH) was employed as a pre-polymer. The photoresist was polymerized by exposure to a 780 nm laser focused by a 63x objective (NA=1.48, Carl Zeiss) directly in contact with the photoresist. Slice by slice, small volumes (400 μm x 400 μm x 100 μm) were exposed using galvanic mirrors and stitched together by moving a mechanical stage. The laser power used was 30 mW. The unpolymerized resist was removed in a bath with PGMA for 15 minutes and rinsed with 2-propanol. The fabrication and the cleaning of the molds were performed under yellow light (λ= 577-597 nm) to prevent spontaneous polymerization. The molds were salinized with trichloro(1H,1H,2H,2H-perfluorooctyl) silane under vacuum for 2 hours to prevent irreversible sticking of PDMS on the surface. A PDMS pre-polymer mixture was prepared from a 1:10 elastomer: crosslinker mixture by weight. It was degassed for 30 minutes under vacuum, poured onto the molds, and further degassed for 2 hours to eliminate any bubbles present, which could affect the shape of the microstructures. Then, the microstructures were cured with temperature for 2 hours at 100ºC for polymerization. Finally, the cured structures were peeled from the molds. Before being used for lipid coating, the microstructures were washed for 10 minutes by sequential sonication in acetone, ethanol, and MilliQ water. The microstructures were dried in an oven and treated with UV/ozone for one hour to fully hydrophilize the surface. The resulting PDMS microstructures were used immediately for the lipid coating to avoid exponential hydrophilization of PDMS.

**Membrane formation:** Membranes were formed by deposition of small unilamellar vesicles (SUVs) obtained via extrusion at high pressure. A mixture of lipids (500 μg) in chloroform (94.8% DOPC, 5% DOPE-PEG (2000), and 0.2% DOPE-Rhodamine, mole ratio) was prepared. The chloroform was evaporated in a vacuum chamber for 2 hours. The lipids were dispersed in HEPES buffer and assembled in multilamellar vesicles during 30 minutes of vortexing. The solution was extruded 21 times with a mini-extruder (Avanti Polar Lipids) equipped with two 250 μL gas-tight syringes (Hamilton), two drain discs, and one nucleopore track-etch membrane with pores of 0.05 μm diameter (Whatman). 50 μL of SUV solution diluted in 1 mL of HEPES buffer was slowly flushed into a flow cell containing the PDMS microstructures and incubated for 1 hour. Finally, the flow cell was flushed with HEPES buffer to remove excess SUVs. Information about the membrane characterization and FRAP measurements can be found in the **Supporting Information Section I-IV.**